\documentclass[twocolumn,amssymb,amsmath,showkeys,aps,prb,showpacs,footinbib]{revtex4} 
\usepackage{graphicx}

\begin{document}
\title{Vortex behavior near a spin vacancy in 2D XY-magnets}
\author{A.R.Pereira, L.A.S.M\'ol}
\address{Departamento de F\'{\i}sica,Universidade Federal de
Vi\c{c}osa,36571-000,
Vi\c{c}osa, Minas Gerais, Brazil}
\author{S.A.Leonel, P.Z.Coura}
\address{Departamento de F\'{\i}sica ICE, Universidade 
Federal de
 Juiz de Fora, Juiz de Fora, CEP 36036-330,Minas Gerais, 
Brazil}
\author{B.V.Costa}
\address{Departamento de F\'{\i}sica, ICEX, Universidade 
Federal de Minas
Gerais,
Caixa Postal 702, CEP 30123-970, Belo Horizonte, Minas 
Gerais, Brazil}
\begin{abstract}
The dynamical behavior of anisotropic two dimensional
Heisenberg models is still a matter of controversy.
The existence of a central peak at all temperatures
and a rich structure of magnon peaks are not yet
understood. It seems that the central peaks are
related, in some way, to structures like vortices.
In order to contribute to the discussion of the
dynamical behavior of the model
we use Monte Carlo and spin dynamics
simulations as well analytical calculations
to study the behavior of vortices in the presence of
nonmagnetic impurities. Our simulations show that
vortices are attracted and trapped by the impurities.
Using this result we show that if we suppose that 
vortices are not very much disturbed by the
presence of the impurities, then they work as
an attractive potential to the vortices explaining
the observed behavior in our simulations.
\end{abstract}
\maketitle
The anisotropic Heisenberg model (AHM) in two dimensions has
received a lot of attention recently. Such attention
is grounded in the fact that a large variety of models
may be mapped in the AHM. It is an interesting model
because it can support topological
excitations, like solitons and vortices, which are present
in several important phenomena.
Topological defects are present in condensed matter systems
such as superconductors, liquid crystals, superfluids, 
magnetic
materials and several others.
The knowledge of how such structures behave
is essential for technological
applications and an important condition for the 
understanding of
many physical questions.
Of great importance is the dynamical behavior of
vortices in ordered structures.
Knowing how vortices are pinned in superconductors is
essential for applications such as magnetic levitation,
improving magnetic resonance imaging devised for medical
diagnosis and many others.
Topological defects can be the signature
left behind by the cosmological phase transitions,
which occur while the universe expands and cools.
Beside all of that,
the study of the dynamics of the AHM model is interesting by
itself. There are several questions not yet responded about
the model. For example,
the origin of the central peak in the
dynamical structure factor observed in experiments and
simulations is the source of several controversial
interpretations. Its origin may be due to vortex
motion\cite{FG,MG,PIRES}. However this point is still
controversial\cite{EVA1,EVA2}.
The existence of some kind of impurity in the system may
affect its properties in several ways. For example, solitons 
near a
nonmagnetic  impurity in 2D antiferromagnets cause 
observable
effects in EPR experiments\cite{KSC,ZKK}. 
Lattice defects such as impurities
and dislocations play a crucial role in disrupting order in
solids.

\begin{figure}[h]
\includegraphics[height=8cm,width=5cm,angle=-90]{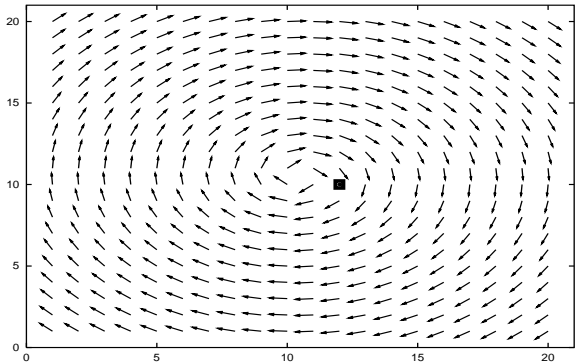}
\caption{\label{DM10}
A typical initial configuration. The square represents
the spinless site.}
\end{figure}

\begin{figure}[h]
\includegraphics[height=8cm,width=5cm,angle=-90]{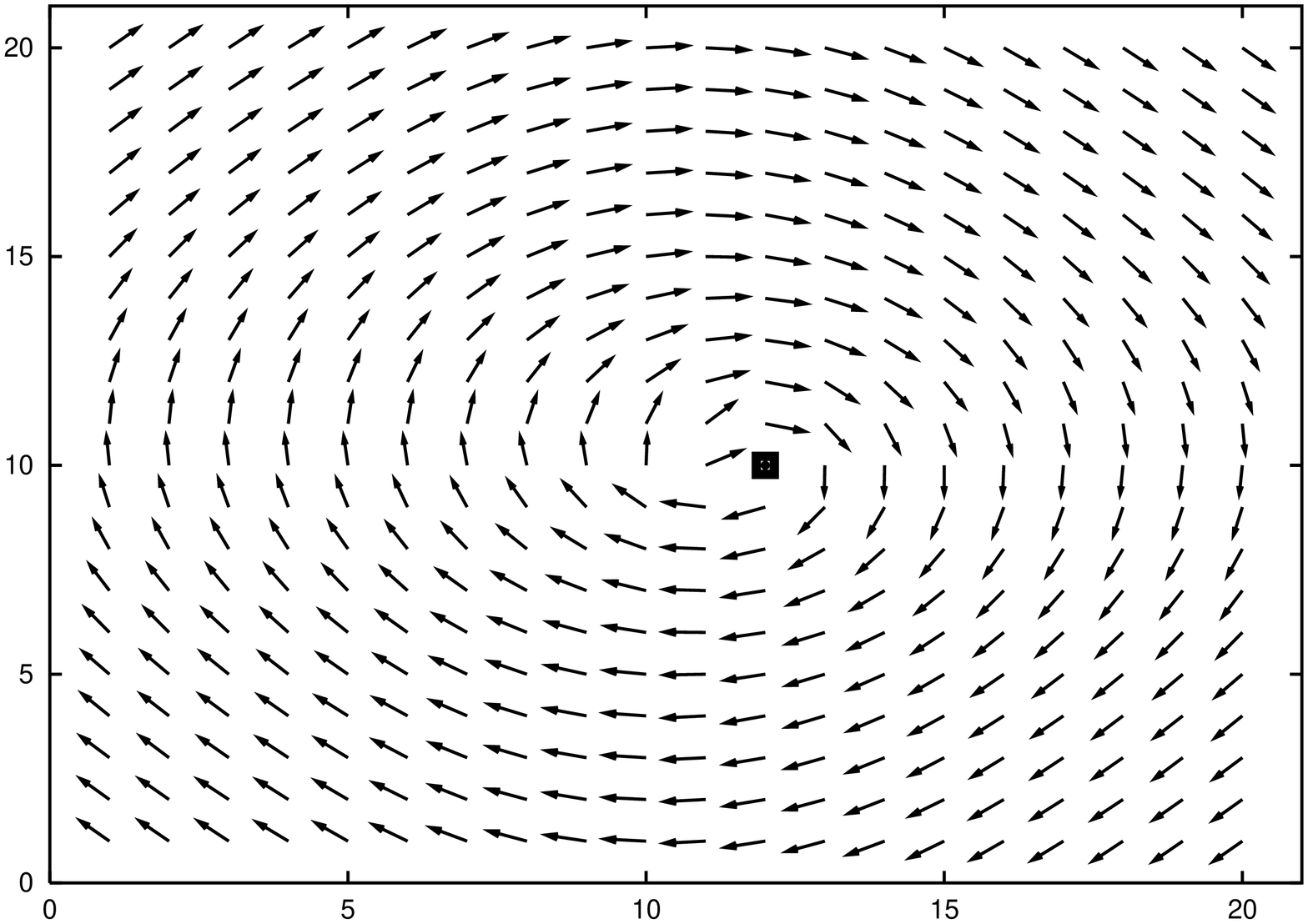}
\caption{\label{DM70}Configuration after 70 time steps.}
\includegraphics[height=8cm,width=5cm,angle=-90]{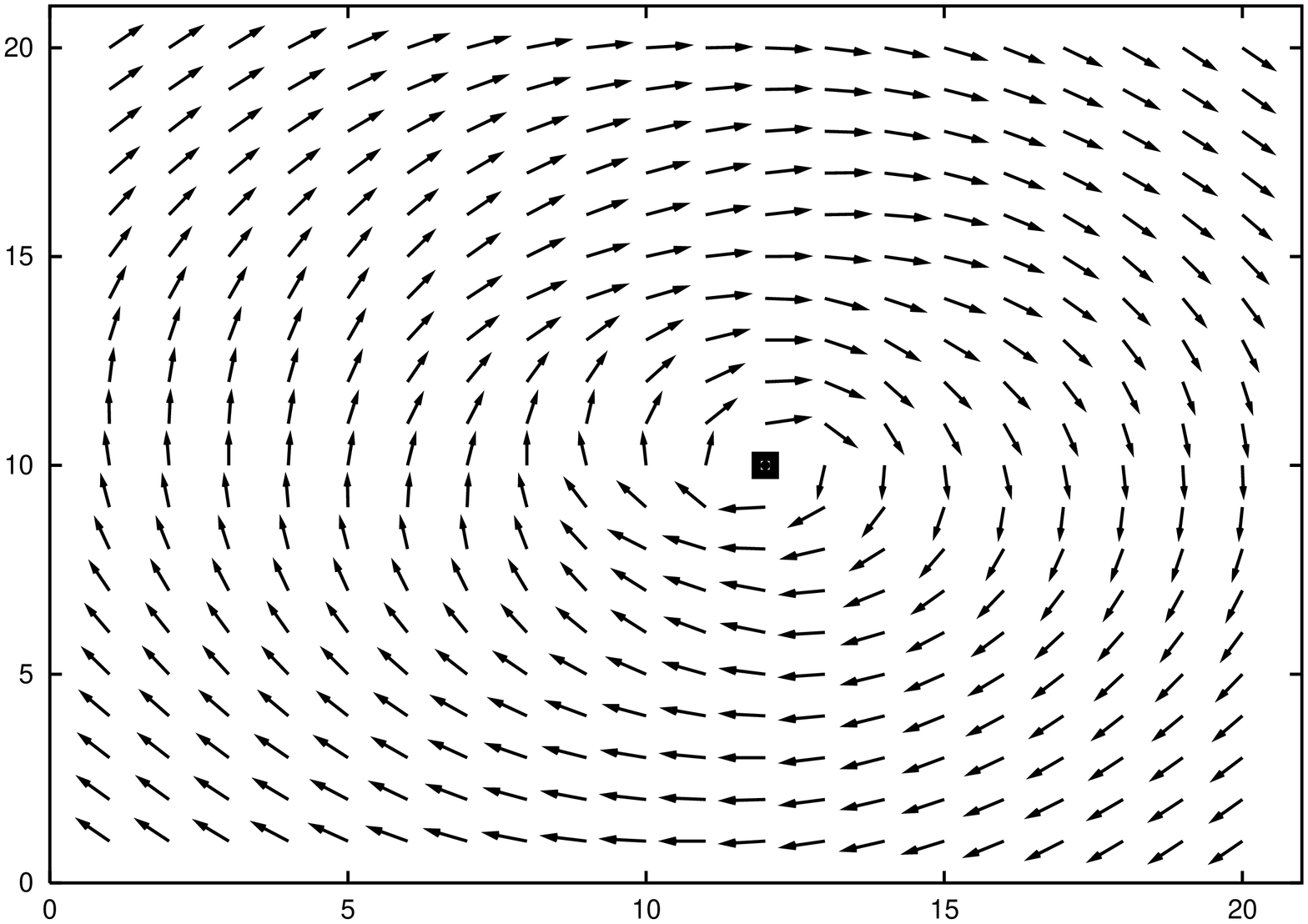}
\caption{\label{DM100}Configuration after 150 time steps.}
\end{figure}

The main task of this letter is to consider the
dynamical effects caused by impurities on
topological defects. We will consider the
classical two-dimensional magnetic (XY) model described by
$H=-J\sum_{m,n}(S^{x}_{m}S^{x}_{n}+S^{y}_{m}S^{y}_{n})$.
Here $J$ is a coupling constant,
$S^{\alpha}$ ($\alpha = x,y,z$)
are the components of the classical spin vector
$\vec S=(S^{x},S^{y},S^{z})$
and the summation is
over nearest-neighbor in a square lattice.
The spin field can be parametrized in terms of spherical
angles as $\vec S=(cos\theta cos\phi,cos\theta
sin\phi, sin\theta)$, where we took $|\vec S | = 1$.
The continuum version of the XY Hamiltonian
can be written as
\begin{eqnarray}
\label{1}
\lefteqn{ H = \frac{J}{2}\int d^{2}r 
\Big[[\frac{m^{2}(\nabla
m)^{2}}{1-m^{2}}+{}}
\\
& & {}(1-m^{2})(\nabla
\phi)^{2}+
\frac{4}{a^{2}}m^{2}\Big].
\nonumber
\end{eqnarray}
where $m=sin\theta$ and $a$ is the lattice constant.
The spin dynamics is given by the equations of motion, $\dot
m= \delta h/\delta \phi$, $\dot \phi=-\delta h/\delta m$, 
where
$h$ is the Hamiltonian density from Eq.(\ref{1}).
The introduction of nonmagnetic impurities
modify the Hamiltonian in the following way\cite{MOL,SA}.
As the model considers only spin interactions up to
nearest-neighbors, to take into account the
presence of a nonmagnetic impurity we introduce a circular
hole with radius of the order of the lattice spacing,
inside of which the Hamiltonian
density vanishes. The Hamiltonian, Eq. (\ref{1}),
becomes\cite{MOL}
\begin{eqnarray}
\label{2} 
\lefteqn {H = \frac{J}{2}\int
d^{2}r \Big [\frac{m^{2}(\nabla
 m)^{2}}{1-m^{2}}+{}}
\\
& & {}(1-m^{2})(\nabla
\phi)^{2}+
\frac{4}{a^{2}}m^{2}\Big ]U(\vec r),
\nonumber 
\end{eqnarray}
with $U(\vec r)=1$ if
$\mid \vec r - \vec r_{0}\mid  \geq a$ and
$U(\vec r)=0$ if $\mid \vec r - \vec r_{0}\mid < a$,
for a nonmagnetic impurity at $\vec r_{0}$.
Early works\cite{MOL,SA} have found a repulsive force
between the vortex and the nonmagnetic vacancy.
The principal ingredient in the calculations
was to suppose that the spin vacancy defforms
considerably the vortex structure.
However, although the assumption of vortex deformation
near the impurity seems reasonable, it needs a convincing
demonstration.
We discuss this question by means
of numerical simulations, which will
be used not only to check the validity
of the approximations done,
but also as a guide to give us a clue on
which assumptions can be made in a
safe way. We have performed spin dynamical simulations
on a $L=20$ square lattice to observe the behavior of
a single vortex initially located at the center of
the system, in the presence of a spin vacancy
located two sites
away from the center of the system.
In order to introduce a single vortex into the
system, we have imposed diagonally antiperiodic
boundary conditions\cite{HIKA}
\begin{eqnarray}
\label{3}
\vec S_{L+1,y}&=&-\vec S _{1,L-y+1}~,~\vec S_{0y}=-\vec S 
_{L,L-y+1}
\nonumber\\
\vec S_{x,L+1}&=&-\vec S_{L-x+1,1}~,~\vec S_{x,0}=-\vec 
S_{L-x+1,L},
\end{eqnarray}
for all $1\leq x,y \leq L$.
The discrete equation of motion for each spin is \cite{EL}
\begin{equation}
\label{4}
{d \vec S_{m}\over dx} = \vec S_{m} \times \vec H_{ef} 
\end{equation}
where
\begin{equation}
\label{5}
\vec H_{ef}= -J\sum_{n} (S^{x}_{n}\hat e_{x} + S^{y}_{n}\hat 
e_{y}) 
\end{equation}
and $\hat e_{x}$ and $\hat e_{y}$ are unit vectors in the
$x$ and $y$ directions,
respectively. The equations of motion were
integrated numerically forward in time
using a vectorized fourth-order Runge-Kutta scheme with a
time step of $0.04 J^{-1}$. The initial structure is shown
in Fig.\ref{DM10}.
In contrast to the analytical results presented in Ref.\cite{MOL}, the
simulation results show that the vortex structure does not change
appreciably near the spin vacancy.
The vortex center
moves toward the spinless site, indicating
an  effective attractive
potential of interaction between the vortex
and the vacancy, as shown in
Figs.\ref{DM70} and \ref{DM100}.
In Fig.\ref{DM70} we show the
configuration after 70 time steps and
Fig.\ref{DM100} after 150 time steps.
We notice that after 150 time steps
the position of the vortex center reaches the
equilibrium with its center at the spinless site.
The key point here is the property of
non-deformation of the vortex configuration.
The attractive potential could also be obtained from
our theoretical model Hamiltonian (\ref{2}) by
assuming  that the vortex configuration is not
deformed due to the presence of an impurity.
Substituting both the non-deformed, static and planar vortex
solution $\phi_{0}=arctan(y/x)$ located at the
origin and a nonmagnetic impurity placed
at the site $\vec r_{0}$ in Hamiltonian  (\ref{2}) we obtain
\begin{eqnarray}
\label{6}
\lefteqn {H_{I}=\frac{J}{2}\int (\nabla \phi_{0})^{2}U(\vec 
r)d^{2}x {}}
\\
& & {}=\frac{J}{2}\Big[\int (\nabla \phi_{0})^{2}d^{2}x-
\int_{A(\vec
r_{0})}(\nabla \phi_{0})^{2}dA(\vec r_{0})\Big],
\nonumber 
\end{eqnarray}
where $A(\vec r_{0})$ is the area of the hole with radius 
$a$
around the point $\vec r_{0}$ that represents the spinless 
site.
The integrals are easily calculated leading to
\begin{equation}
\label{7}
E_{I}=E_{\nu}+\frac{\pi 
J}{2}ln\left(1-\frac{a^2}{r_{0}^{2}}\right),
\end{equation}
where $E_{I}(\vec r_{0})$ and $E_{\nu}=\pi J ln(L/a_{0})$ 
are the vortex energies in the presence and absence of the
nonmagnetic impurity respectively for a system of size $L$.
The constant $a_{0}=0.24a $ \cite{GM} present in $E_{\nu}$,
is some suitable short-distance cutoff to avoid spurious
divergences due to the fact that the vortex center is a
singularity in the continuum limit. Its appropriated value
for the square lattice was determined numerically
in Ref.\cite{GM}.
Of course, Eq.(\ref{7}) is not valid in the limit
$r_{0}\rightarrow 0$, where the  continuum approximation
breaks down. In this limit there will
have an intersection between the circular hole of the
impurity and  the structure of the vortex core. A point of
the circumference  of the impurity circle would meet the
center of the vortex core 
lowering the vortex energy as $\vec r_{0}$ decreases. Then, from Eq.(\ref{7})
the effective potential experienced by the two defects can be written as
\begin{eqnarray}
\label{8} 
\lefteqn {V_{eff}(r_{0})=E_{I}(r_{0})-E_{\nu}{}}
\\
& & {}=\frac{\pi
J}{2}ln\left(1-\frac{a^{2}}{r_{0}^{2}}\right) 
\mbox{, for $r_{0}>a$,} 
\nonumber 
\end{eqnarray}
which is attractive. For large distances of separation
$(r_{0}>>a)$, the Eq.(\ref{8}) can be approximated by
$V_{eff}(r_{0})\approx -(\pi Ja^{2}/2)(1/r_{0}^{2})$. To 
know the
behavior of the potential in the region inside the core
$(r_{0}\leq a)$, we first put the hole exactly at the
vortex
center, obtaining $E_{I}=\pi Jln(L/a)$, which leads to
$V_{eff}(0)=E_{I}(0)-E_{\nu}=\pi Jln(0.24)=-4.48J$. Then, 
using this result, we modify Eq.(\ref{8}) to get a
complete function $V_{eff}(r_{0})$, introducing a small 
constant b as follows
\begin{equation}
\label{9}
V_{eff}(r_{0})=\frac{\pi J}{2}
ln\left(1-\frac{a^{2}}{r_{0}^{2}+b^{2}}\right),
\end{equation}
with the condition
\begin{equation}
\label{10}
V_{eff}(0)=\frac{\pi J}{2}
ln\left(1-\frac{a^{2}}{b^{2}}\right)=\pi Jln(0.24).
\end{equation}
This artifact leads to an expression valid in all space that
reproduces correctly the asymptotic limits for $b=1.03a$.
The potential (Eq. \ref{9}) indicates that vortex can be 
trapped by a nonmagnetic impurity. The
attractive potential experienced by the vortex  can be
written as
\begin{equation}
\label{11}
V(X)=\frac{\pi 
J}{2}ln\left(1-\frac{a^{2}}{X^{2}+b^{2}}\right),
\end{equation}
where $X$ is the distance between vortex center and the 
spinless site.
\begin{figure}[h]
\label{potential}
\includegraphics[height=5cm,width=8cm]{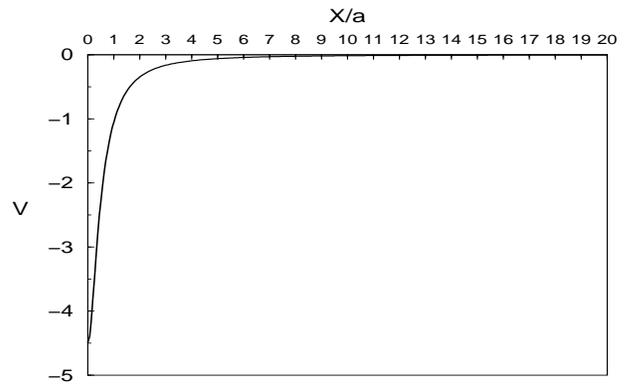}
\caption{\label{potencial}Attractive potencial as a 
function of $X/a$}
\end{figure}
It has a minimum at $X=0$ and decreases rapidly
for $X> 2a$ (See Fig.\ref{potencial} ).

\begin{figure}[h]
\includegraphics[height=8cm,width=5cm,angle=-90]{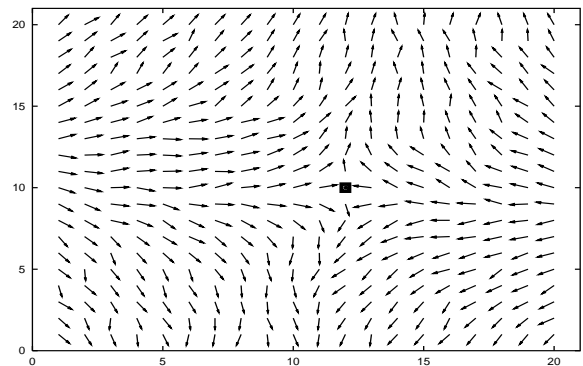}
\caption{\label{MCA7}Anti-vortex configuration after 
$2\times 10^{5}$ MC
steps. The impurity is located at site (12,10).}
\end{figure}

\begin{figure}[h]
\includegraphics[height=8cm,width=5cm,angle=-90]{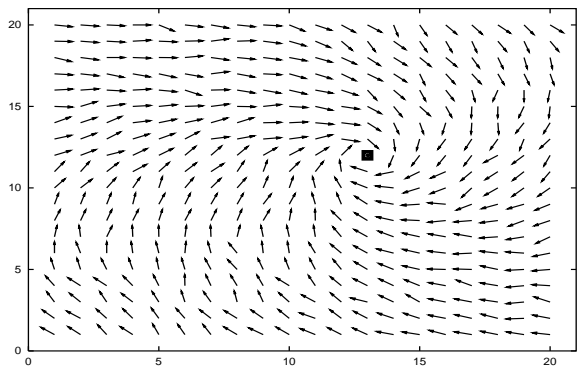}
\caption{\label{MCV7}Vortex configuration after $2\times 
10^{5}$ MC
steps. The impurity is located at site (13,12).}
\end{figure}

In order to observe the lowest energy configurations at low
temperatures in a
classical two-dimensional system described by the $XY$ 
model, with a
nonmagnetic impurity, we performed Monte Carlo (MC) 
simulations. The
simulations were done on a $L=20$ square lattices at 
temperature $T=0.1$ in
units of $J/k_{B}$ ($k_{B}$ is the Boltzmann constant), using a 
standart Metropolis
algorithm \cite{MET} with diagonally antiperiodic boundary 
conditions
(Eq.\ref{3}) and random initial configurations. Using these 
boundary
conditions, at low temperature, the equilibrium 
configurations can have a
single vortex or a single anti-vortex. We have observed that 
after $10^{5}$ MC
steps the equilibrium configurations was reached. 
Fig.\ref{MCA7} shows a
single anti-vortex equilibrium configuration after $2\times 
10^{5}$ MC steps
with the nonmagnetic impurity located at site (12,10). Fig. 
\ref{MCV7} shows a
single vortex equilibrium configuration after $2\times 
10^{5}$ MC steps with
the nonmagnetic impurity located at site (13,12). We have 
observed that, in
both cases, in the lowest energy configurations, the 
anti-vortex and vortex
centers are located at the nonmagnetic impurity, which is in 
agreement with
Eq.(\ref{11}).

In summary, using Monte Carlo and
molecular dynamics simulations, we observed that a
vortex structure put close to a non magnetic
site in a square lattice
does not change appreciably its geometry.
As the system evolutes in time the
vortex center moves toward the spinless site.
By using a continuum approach
we have  modeled the interaction as
an attractive potencial which has a
minimum for the vortex center located at spinless site, as 
observed in MC simulations.
The study of the dynamics of diluted models
can help us to understand the origin of the central
peak found in early simulations. The central peak 
is believed to be due to vortex motion.
If that is the important contribution,
the presence of non magnetic sites will diminish the
high of the central peak due to a trapping of
vortices at that non magnetic sites.

This work was partially supported by CNPq and 
FAPEMIG(Brazilian agencies).
Numerical work was done at the Laborat\'orio de 
Computa\c{c}\~ao e
Simula\c{c}\~ao do
Departamento de F\'{\i}sica da UFJF.


\begin{thebibliography}{100}
\bibitem{FG}F.G.Mertens, A.R.Bishop,G.M.Wysin, and 
C.Kawabata, Phys.Rev.B
{\bf30},591(1989).
\bibitem{MG}M.E.Gouv\^ea,G.M.Wysin.A.R.Bishop, and 
F.G.Mertens, Phys.Rev.B
{\bf 89}, 11840(1989).
\bibitem{PIRES}A.R.Pereira, A.S.T.Pires, M.E.Gouv\^ea and
B.V.Costa, Z. Phys. B: Condens. Matter {\bf 89},109(1992).
\bibitem{EVA1} J.E.R. Costa and B.V. Costa,
Phys.Rev.B {\bf 54}, 994(1996).
\bibitem{EVA2}B.V.Costa, J.E.R. Costa and D. P. Landau,
J. Appl. Phys. {\bf 81}, 5746(1997).
\bibitem{KSC}K.Subbaraman,C.E.Zaspel, and
K.Drumheller,Phys.Rev.Lett.80,2201(1998).
\bibitem{ZKK}C.E.Zaspel, K.Drumheller, and
K.Subbaraman,Phys.Stat.Sol.(a)189,1029,(2002).
\bibitem{MOL}L.A.S.M\'ol, A.R.Pereira, and A.S.T.Pires,
Phys.Rev.B {\bf 66}, 052415(2002).
\bibitem{SA}S.A.Leonel, P.Z.Coura, A.R.Pereira, L.A.S.M\'ol
and B.V.Costa, Phys.Rev.B {\bf 67}, 104426(2003).
\bibitem{HIKA}H.Kawamura and M.Kikuchi, Phys. Rev.B {\bf
47}, 1134(1993). 
\bibitem{EL}H.G.Evertz and D.P.Landau, Phys. Rev.B {\bf 54},
12302(1996).
\bibitem{GM}G.M.Wysin, Phys.Rev.B {\bf 54}, 15156(1996).
\bibitem{MET}N.Metropolis,A.W.Rosenbluth,M.N.Rosenbluth,A.H.Teller
and E.Teller,J.Chem.Phys.21,1087(1953) and for a 
review,see,e.g Monte Carlo Methods in Statistical Physics, edited by
K.Binder(Springer, New York 1979). \end{thebibliography}
\end{document}